\documentclass[aps,pra,10pt,superscriptaddress,numerical,floatfix,twocolumn]{revtex4-2}

\usepackage{bm}
\usepackage{braket}
\usepackage{color}
\usepackage{dcolumn}
\usepackage{epsfig}
\usepackage[T1]{fontenc}
\usepackage{graphicx}
\usepackage[colorlinks]{hyperref}
\AtBeginDocument{
	\hypersetup{
		urlcolor=black,
		citecolor=green,
		linkcolor=blue,
	}%
}
\usepackage[latin9]{inputenc}
\usepackage{mathtools}
\usepackage{subcaption}
\usepackage{xcolor}
\usepackage{cleveref}
\usepackage{amssymb}
\usepackage{amsthm}
\usepackage{amsmath}

\def\app#1#2{%
	\mathrel{%
		\setbox0=\hbox{$#1\sim$}%
		\setbox2=\hbox{%
			\rlap{\hbox{$#1\propto$}}%
			\lower1.2\ht0\box0%
		}%
		\raise0.25\ht2\box2%
	}%
}

\begin{document}
	\title{Constructing Nano-Object Quantum Superpositions with a Stern-Gerlach Interferometer}


\newcommand{\affone}{Department of Physics and Astronomy, University College London, Gower Street, WC1E 6BT London, United Kingdom.}
\newcommand{\afftwo}{Van Swinderen Institute, University of Groningen, 9747 AG Groningen, The Netherlands.}
\newcommand{\affthree}{Department of Physics, Ben-Gurion University of the Negev, Be'er Sheva 84105, Israel.}

\author{Ryan J. Marshman}
\affiliation{\affone}

\author{Anupam Mazumdar}
\affiliation{\afftwo}

\author{Ron Folman}
\affiliation{\affthree}

\author{Sougato Bose}
\affiliation{\affone}
   
\date{\today}

\begin{abstract}
	{Probing quantum mechanics,  quantum aspects of general relativity, along with the sensing and the constraining of classical gravity can all be enabled by unprecedented spatial sizes of superpositions of massive objects. In this paper, we show that there is a feasible setup sourced by realisable magnetic field gradients ${\cal O}(10-100)$~Tm$^{-1}$ to construct a large spatial superposition of ${\cal O}(10^{-4}-10^{-8})$~m  for ${\cal O}(10^{-17}-10^{-14})$~kg masses within a time of up to $0.1-10$~seconds. The scale of superpositions is unrestricted as long as quantum coherence can be maintained for a required amount of time.}

\end{abstract}

\maketitle

\section{Introduction}
At a microscopic level, three of the known forces of nature, electromagnetic (EM), weak and strong, obey the principles of local quantum field theory (QFT) \cite{Weinberg:1995mt}. However, there is no experimental proof yet of {\em how} the gravitational interaction is being mediated. Recently, a tabletop experiment has been suggested to explore the quantum origin of gravity in a tabletop experiment~\cite{Bose:2017nin, QNG, Marletto2017}. The protocol crucially relies on the interaction of quantum matter with quantum gravity, leading to the generation of entanglement between spins embedded in two non-relativistic test masses~\cite{QNG}. The spin entanglement witness will prove the graviton's quantum-ness as a mediator of the force between the two masses (quantumness of the linearized metric fluctuations around a Minkowski background), and will further test the nature of the gravitational interaction at microscopic distances~\cite{QNG,Biswas:2011ar}. However, there remains a demanding requirement: {\em quantum spatial superpositions of distinct localized states of neutral mesoscopic masses} $m \sim 10^{-15}$~kg over spatial separations of $\Delta x \sim 10$~$\mu$m \cite{vandeKamp:2020rqh}, far beyond the scales achieved to date (e.g., macromolecules $m\sim 10^{-22}$~kg over $\Delta x \sim 0.25$~$\mu$m, or atoms $m\sim 10^{-25}$~kg over $\Delta x \sim 0.5$~m)~\cite{Arndt:1999kyb,arndt:2014testing}.
Such superpositions have also been shown to be of practical value in sensing of weak forces, curvature \cite{torovs2021relative}, frame-dragging, and even a tabletop detection of low-frequency gravitational waves \cite{MIMAC2018}.

Beyond the usage in sensing quantum and classical gravity, upgrading the mass $m$ and the superposition $\Delta x$, naturally stretches the boundaries of the validity of QM, which in itself is a worthy goal \cite{leggett2002testing, arndt:2014testing}. However, there is a gap in the literature at the moment as far as a realistic scheme for achieving superposition sizes $\Delta x > 1$~$\mu$m. While there are quite a few schemes for a lower $\Delta x$, or $m$, investigated at various levels of detail \cite{bose1999scheme, armour2002entanglement, marshall2003towards, sekatski2014macroscopic, romero2010toward, romero2011large, khalili2010preparing, scala2013matter, PhysRevLett.117.143003, bateman2014near, yin2013large,pino2018chip, clarke2018growingPublished, ringbauer2018generation, khosla2018displacemon, kaltenbaek2012MAQRO, romero2017coherent, Pedernales2020, hogan2008light, margalit2021realization, al2018optomechanical}, and while these may suffice to falsify various purported modifications of QM \cite{bassi2013models, kovachy2015quantum, murata2015review}, the only two predictions stemming from the straightforward application of QM (local QFT to be more precise \cite{QNG}) which would give us non-trivial information (e.g. sensing the quantum nature of gravity \cite{Bose:2017nin} and sensing extremely weak classical gravity \cite{MIMAC2018}), necessarily seem to require $\Delta x \gtrsim 1$~$\mu$m. 

Rudimentary arguments on how to achieve such superpositions using a Stern-Gerlach Interferometer (SGI) in high magnetic field gradients were presented in \cite{Bose:2017nin}, building on earlier ideas for smaller $m, \Delta x$ superpositions \cite{PhysRevLett.117.143003,wan2017quantum}. A feasibility study building on atomic experiments was recently performed, showing that SGI for massive objects is indeed possible~\cite{margalit2021realization}. However,  beyond a simple scaling of mass, little work has been done \cite{Pedernales2020} on exploring what new issues will arise when these techniques are pushed beyond the atomic level.

In this paper, we go further in several respects by incorporating some crucial aspects missed in previous treatments while making it simultaneously less demanding in certain aspects, and confirm the true potential for the spatial splitting of massive objects (namely nanocrystals) achieved by SGI.  The proposal outlined here requires only moderate magnetic gradients, significantly less than what was originally considered necessary\cite{Bose:2017nin} as shown in Figure \ref{fig:superposition with time and mass}. We will be describing a 1D longitudinal interferometer, which avoids the problems noted for 2D interferometers~\cite{paraniak2021quantum}. We take into account gradients in other directions, as demanded by Maxwell equations. We take into account the effects of the induced diamagnetism within the interferometric mass as while this is not of concern with atomic Stern-Gerlach interferometry, it becomes a dominant effect once larger masses are considered. We consider a magnetic source which enables constant gradients over a large volume. In contrast to the previous works \cite{scala2013matter, PhysRevLett.117.143003, Pedernales2020}, we have a scheme without a low magnetic field in any region, so that to avoid the historically well known phenomenon of Majorana spin flips \cite{majorana1932atomi, inguscio2006majorana}. In this way we have {\em combined} well studied problems in particle trapping (Majorana spin flips),  and practical realizations of Stern-Gerlach effect to show that a full-loop interferometer \cite{margalit2021realization}, where wavepackets are spatially spilt in a spin dependent manner and then brought back to overlap both in position and momentum to complete the interferometry,  is still possible with nanocrystals.

The above results are {\em simultaneously} achieved by a significant modification of standard SGI by changing the initial conditions and incorporating a gradient-free spatial region in which the diamagnetic force does not act, and yet, the superposition continues to grow in view of a momentum difference. However, we show what appears to be an unavoidable price to pay, that  is a linear growth of a relative phase between the superposed components in time. These features allow us to present in detail a feasible interferometric accelerator for micro-scale masses, and enable the creation of a massive Schr\"odinger's cat.

\begin{figure}
	\centering
	\includegraphics[width=0.95\linewidth]{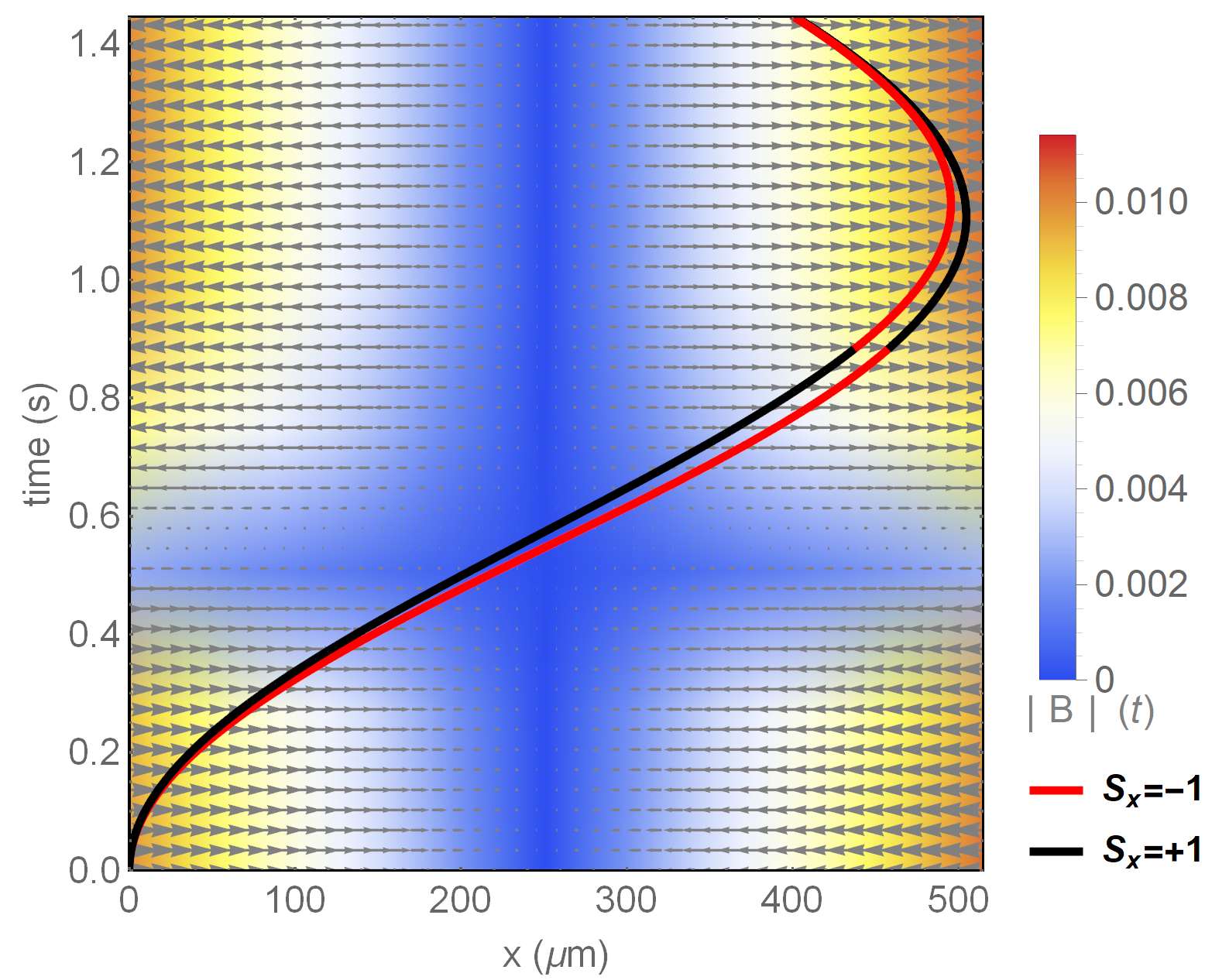}
	\caption{Example interferometer path trajectories through the varying magnetic fields used. The magnetic field transition time here is accentuated for readability. Note that the internal spin states are reversed at time $t\approx0.9$ s to ensure the interferometer is closed. \label{fig:Trajectories and Bfield}}
\end{figure}

Previous experimental configurations consider a magnetic field which originates in a single current-carrying wire or a permanent magnet, whereby the field goes as $\vec{B}\propto1/r$, where $r$ is the distance from the source. The magnetic field can then be linearised in a small region within which it has an approximately constant gradient. However, a large splitting will require long evolution times, and as the distance quickly increases between the particle and the wire, significant magnetic gradients are no longer available. We will therefore consider a configuration which enables a constant gradient over a large region (e.g. quadrupole field from coils in an anti-Helmholtz configuration). A typical resultant trajectory of such a potential is shown in Figure~\ref{fig:Trajectories and Bfield}. 

\section{Model}
We assume a host nano-crystal with a single spin embedded in it, and concentrate, in this work, only on the translational motion of this crystal under a magnetic field gradient. To explicitly model the spin and the host crystal material, we will assume a nitrogen-vacancy (NV) centre spin in a diamond nano/micro-crystal, although this analysis holds for all materials with similar diamagnetic susceptibilities and generic electronic spins.  We consider the mass to be in a {\it free fall} along the $z$-axis. We therefore do not treat its motion along the $z$ axis explicitly as here the $z$ motion commutes with the motion along the other two axes.  Moreover, $z$ motion would be absent in a drop-tower experiment, for example, which may be necessary in such experiments because of their long durations of $\sim 1$ s, and to mitigate gravitational jitters \cite{torovs2021relative}.  The Hamiltonian of the system is given by~\cite{Loubser_1978,Pedernales2020}
\begin{equation}
	H=\frac{\hat{P}^2_x+\hat{P}^2_y+\hat{P}^2_z}{2m} + \hbar D\hat{S}^2_{x}-\frac{\chi_m m}{2\mu_0}\vec{B}^2-\mu_B \hat{s}\cdot\vec{B}\,, \label{eq:full hamiltonian} 
\end{equation}
where $m$ is the mass of the diamond, $\chi_m=-6.2\times10^{-9}$ m$^3$kg$^{-1}$ is the mass magnetic susceptibility, $\mu_B$ is the Bohr magneton, $\hat{s}$ is the spin operator, and $D=\left(2\pi\right) \times2.8$~GHz is the NV zero-field splitting.  Note that in writing the above Hamiltonian, we are only concentrating on the {\em translational motion} -- we are assuming that we can engineer a situation so that the torque and the rotational effects of the mass is negligible/decoupled from translation (for possible mechanisms to cool rotation see~\cite{kuhn2017full, schafer2021cooling, rudolph2021theory, van2021sub}). The external magnetic field $\vec{B}$ varies in the $x-y$ plane, and as we will see, with our choice of its profile, couples predominantly to the $x$ motion via the $x$-component of the spin. Under this setting, once the internal spin of the NV centre is initialised in the state $\frac{1}{\sqrt{2}}\left(\left|1\right\rangle+\left|-1\right\rangle\right)$, we will realize a 1-dimensional SGI which, as time progresses, achieves first a maximum wavefunction splitting $\Delta x$ in the $x$ direction, and then, via appropriate spin flips/magnetic gradient changes, its subsequent recombination to complete the interferometry. The second term of the Hamiltonian drops out as a common phase and we will omit this term from our analysis. This also means that our results will be completely general, and hold for other (say spin$-1/2$) dopants in generic crystals. 


There are known predicaments for generating superpositions by using a spin coupled to a magnetic field: (1) The electronic spin will in general experience Majorana spin-flips whenever the magnetic field magnitude becomes small, thus no longer remaining in a well-defined eigenstate as required for the  coherent manipulation of the masses (this is particularly important for our results to be inclusive of generic spin dopant atoms in a nanocrystal; it has been perceived as an important problem for several atomic species \cite{inguscio2006majorana}), and (2) the effect of off-axis magnetic field gradients, which are generally not considered~\cite{Pedernales2020}, however, this \emph {must} be taken into account to satisfy Maxwell's equations. 

First, we will consider the constraints on the magnetic field to address the above problems. Noting that $\vec{\nabla}\cdot\vec{B}=0$ and $\vec{\nabla}\times\vec{B}=0$, as we keep the mass away from the source of the magnetic field. For simplicity, we assume the following profile in the $(x,y)$ plane: $\vec{B}=B_x(x,y) \hat{x} + B_y(x,y) \hat{y}$. 
We take the $x$ axis as the {\it desired superposition direction}, and we require the magnetic field to be linear along $x$ direction, with a constant magnetic field gradient along the $x$ direction to take some constant value $\partial B$. To ensure that we can take the x-axis as our quantisation axis, and that Majorana spin flips are avoided, we require that $\left|B_x(x,y)\right|\gg\left|B_y(x,y)\right|\forall x,y$ in the vicinity of the controlled trajectories. The simplest general form for the magnetic field which will satisfy all the above conditions can be given by: 
\begin{equation}
	\vec{B}(x,y)=\left(B_x(0,0)-\partial B x\right)\hat{x} + \partial B y\hat{y} \,,\label{eq:General linear magnetic field}
\end{equation}
for any fixed value of the magnetic field at the origin $B_x(0,0)$ and magnetic field gradient $\partial B$. Note that for positive $B_x(0,0)$ and $\partial B$ the zero point of the magnetic field is always on the positive $x$ axis. By ensuring that $y\approx0$ during the entire interferometry, we find a suitable definition for the zero magnetic field region which must be avoided of
\begin{equation}
	x\notin\left[\frac{B_x(0,0)-\varepsilon}{\partial B},\frac{B_x(0,0)+\varepsilon}{\partial B}\right]\,, \label{eq:limit on x}
\end{equation}
where $\varepsilon$ is the minimum allowable magnetic field in the $x$ direction. For a sufficiently large $\varepsilon$, we can ascertain that the spin states will always be approximately aligned along the quantisation ($x$) axis and Majorana spin flips are avoided.

The spin state in the $y$ and $z$ basis will experience a rapid Larmor precession, with a frequency set by $\omega_L=-\frac{g e}{2 m_e}|B(x,y)|$, where $g\approx2$ is here the Land\`e g factor, $e$ is the electron charge and $m_e$ is the electron mass. If we desire a minimum Larmor frequency of $\omega_L^{min}$, we can define a minimum allowable magnetic field magnitude:
\begin{equation}
	\varepsilon\sim -\frac{2 m_e \omega_L^{min}}{g e}.
\end{equation}
Therefore, the particle must not enter the region given by Eq.(\ref{eq:limit on x}). 
To achieve this, we will use three linear magnetic field profiles sequentially in time, which are specific solutions of the form of Eq.(\ref{eq:General linear magnetic field}):

\begin{eqnarray}
	\vec{B}\left(x,y\right)=&\left(B_0-\eta x\right)\hat{x}+\eta y\hat{y} \label{eq:Bfield1}\\
	\vec{B}\left(x,y\right)=&B_1\hat{x} \label{eq:Bfield2}\\
	\vec{B}\left(x,y\right)=&-\left(B_0-\eta x\right)\hat{x}-\eta y\hat{y} \label{eq:Bfield3}
\end{eqnarray}
with $B_0,B_1,\eta>0$. We then require $B_1\gg\varepsilon$, and the timing of the switching between the magnetic fields will be done to ensure that the particle never experiences a small magnetic field. That is, as the particle approaches the disallowed region given by Eq. \ref{eq:limit on x} (for $B_x(0,0)=B_0$ and $\partial B=\eta$), the magnetic field is mapped to that given by Eq. \ref{eq:Bfield2} (where now $B_x(0,0)=B_1$ and $\partial B=0$, thus the region specified by Eq. \ref{eq:limit on x} does not occur), and as it leaves that region, the magnetic field is smoothly mapped to Eq. \ref{eq:Bfield3} (where $B_x(0,0)=-B_0$ and $\partial B=-\eta$). The switching function will be modelled by
	$\textrm{Sw}(t,t_{on},t_{off}) =0.5\left(\tanh\left[\delta\left(t-t_{on}\right)\right]+1\right)\times0.5\left(\tanh\left[\delta\left(t_{off}-t\right)\right]+1\right)$
where $\delta$ is the switching frequency parameter, which we ensure to be sufficiently slow such that the magnetic field change is always adiabatic. This is to ensure that it does not complicate the spin dynamics. We consider $\delta=10^{3}$~Hz, which is well below what is required to maintain the adiabaticity conditions $\dot \omega_L/\omega_L^2 \ll 1$ and $\delta\ll\omega_L^{min}$. With this, since the magnetic field is along a fixed direction, it effectively freezes the spin direction (with the other spin components experiencing high-frequency Lamor precession) ensuring the interferometer is {\it effectively} one dimensional.

The motion of the particle can be separated into different stages, largely depending on the form of $\vec{B}$:
\begin{enumerate}
	\item $t<\tau_1$, $\vec{B}$ is given by Eq.\ref{eq:Bfield1}.
	\item $\tau_1\le t<\tau_2$, $\vec{B}$ is switching adiabatically from Eq.\ref{eq:Bfield1} to Eq.\ref{eq:Bfield2}, via the switching function
	$\textrm{Sw}(t,t_{on},t_{off})$.
		\item $\tau_2\le t<\tau_3$, $\vec{B}$ is given by Eq.\ref{eq:Bfield2}.
	\item $\tau_3\le t<\tau_4$, $\vec{B}$ is switching adiabatically from Eq.\ref{eq:Bfield2} to Eq.\ref{eq:Bfield3} via the switching function
	$\textrm{Sw}(t,t_{on},t_{off})$.
	\item $\tau_4\le t$, $\vec{B}$ is given by Eq.\ref{eq:Bfield3}.
	\item $t=\tau_5$, when the spin states are reversed to close the superposition while $\vec{B}$ is still given by Eq.\ref{eq:Bfield3}.
	\item $t=\tau_6$, when the two wavefunctions are brought to overlap in the position and the momentum basis.
\end{enumerate}

Away from the nearly zero-field region, where the magnetic field profile is given by Eq.\ref{eq:Bfield1} or \ref{eq:Bfield3}, we can write the potential energy in a compact form as:
\begin{equation}
	U_{\pm}(\tilde{x}')=-\frac{\chi_m m}{2 \mu_0}\eta^2\tilde{x}'^2 +\mu_{\pm}\frac{\mu_{0}}{\chi_{m}m} \label{eq:SHO for electronic spin}
\end{equation}
where we have taken $y\approx 0$, while  the second term is a constant energy, and $\tilde{x}'=x-C(s_{x})$, where
\begin{equation}
	C(s_{x}=\pm1)= {B_0}/{\eta}\pm\frac{g e \hbar}{2 m_e}\frac{\mu_{0}}{\chi_{m} m \tilde{\eta}(t)}\label{eq:electronic spin solution}\,,
\end{equation}
where $\tilde{\eta}(t)=\eta$ when $t<\tau_1$ and $\tilde{\eta}(t)=-\eta$ when $t>\tau_4$.
Thus the object sees a harmonic potential created by the diamagnetic interaction whose centre is displaced  in the $x$ direction by the spin-magnetic field gradient interaction.  It is clear from the above that the object will roll in different potential wells corresponding to its spin state and thereby develop a momentum difference as they approach the nearly zero field region. It is in this region that we switch the gradient (the harmonic potential) off, and let the object evolve in a magnetic field given by Eq.(\ref{eq:Bfield2}) so that a continually increasing spatial splitting can develop due to the momentum difference of the spin components, while Majorana spin flips are avoided. 
\begin{figure*}
	\begin{subfigure}{0.51\textwidth}
		\centering
		\includegraphics[width=0.99\linewidth]{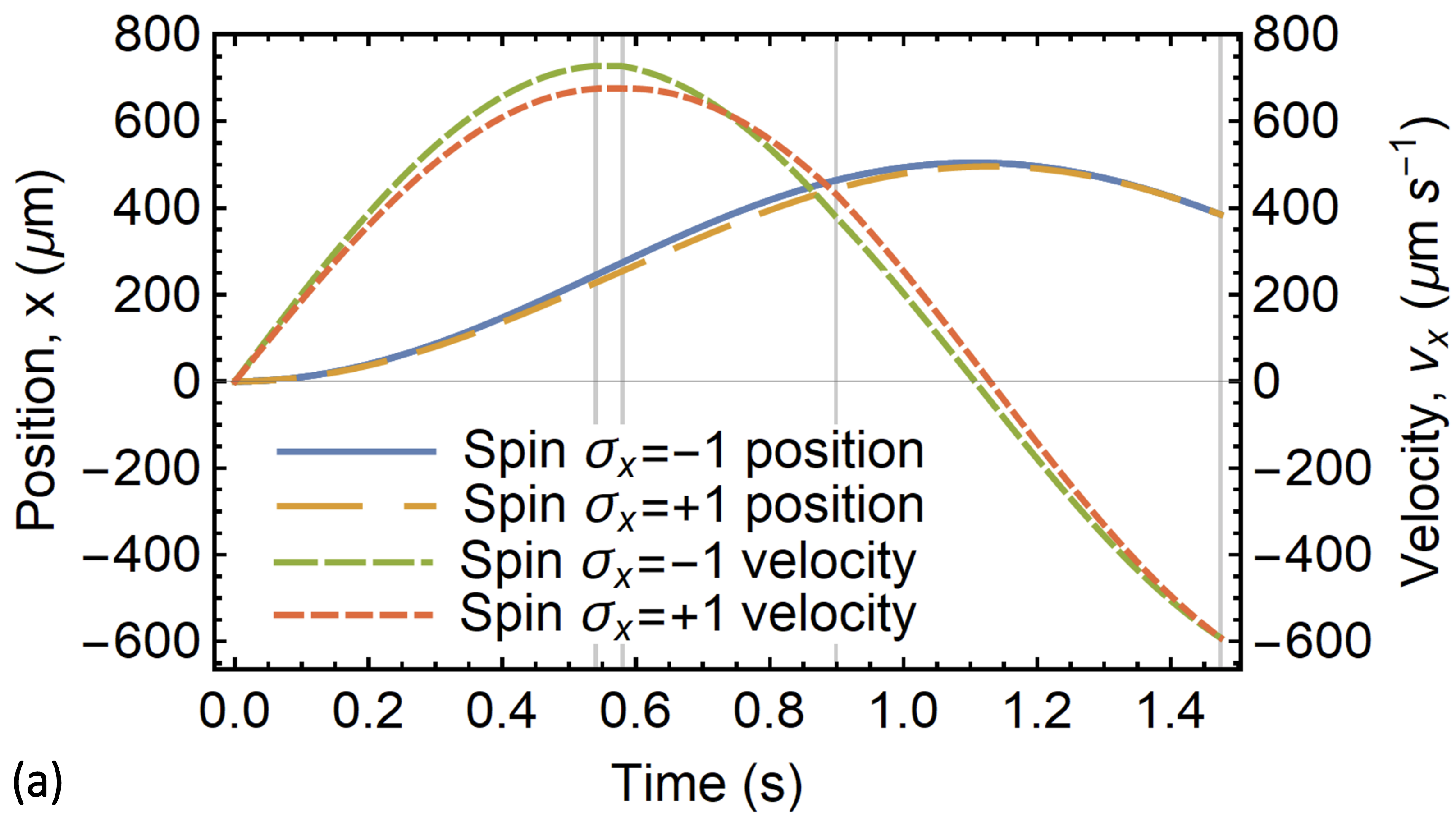}
	\end{subfigure}
	\begin{subfigure}{0.48\textwidth}
		\centering
		\includegraphics[width=0.90\linewidth]{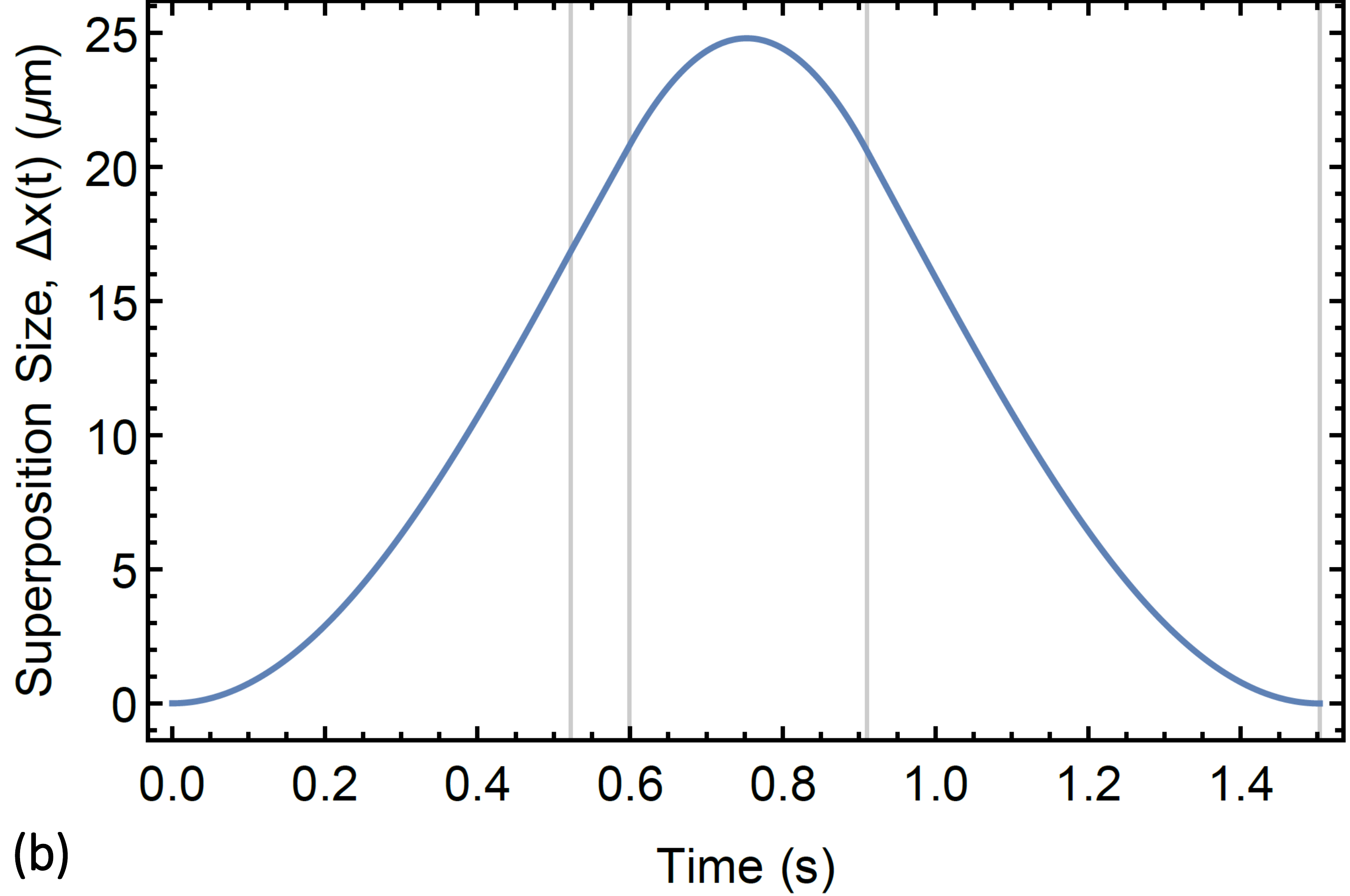}
	\end{subfigure}
	\caption{Figure (a) shows both the paths taken through the interferometer and the corresponding velocity while figure (b) shows the superposition size with time. These plots are the result of the combined numerical and analytical analysis. The vertical grey lines represent times $\tau_1\approx\tau_2$, $\tau_3\approx\tau_4$, $\tau_5$ and $\tau_6$. Both figures are for $B_0=10^{-2}$~T, $\eta=46.483$~Tm$^{-1}$ and $m=10^{-17}$~kg.}
	\label{fig:Path figures}
\end{figure*}
When the masses are in a coherent state of the harmonic potential, the wavefunction will not spread, while when in the constant magnetic field (free motion) it will hardly spread due to the largeness of the mass. Thus, it suffices to model the centre of mass as following a classical trajectory for each interferometric path. Thermal fluctuations in the initial state do not limit the coherence in the final state as they factor out of the motion, see the discussions in \cite{bose1999scheme,PhysRevLett.117.143003}.
To a good approximation, the two interferometric paths can be modelled by $x(t)=A \cos\left(\omega t + \phi\right) + C(s_{x})$, where the frequency of the diamagnetic trap is given by $\omega=(-{\chi_m}/{\mu_0})^{1/2}\eta$, while note that $\chi_{m}<0$ as the mass is diamagnetic.
This leaves the amplitude $A$ and the phase $\phi$ to be determined by requiring that the position and the momentum of each arm of the interferometer are continuous throughout the trajectory. For the ease of computation, we have solved the complete trajectories using an appropriate combination of analytic and numerical solutions. For periods of evolution by a time independent Hamiltonian, analytic solutions are used. However to allow for a more realistic magnetic field switching to be modelled, numerics are used, specifically during $\tau_1\leq t< \tau_2$ and $\tau_3\leq t< \tau_4$. This is discussed in more detail in Appendix \ref{sec:solving the dynamics}. The resulting motion can be seen in Fig.\ref{fig:Path figures}. Note that at $\tau_5\approx 0.9$~s (marked by the vertical lines in Fig.\ref{fig:Path figures}), the spin reversal takes place by firing a rapid microwave pulse to alter the internal spin state.  We can find numerically the final time $\tau_6$ to be 
\begin{equation}
	\tau_6\approx59\times ({1{\rm~Tm^{-1}}}/{\eta})~{\rm~sec}\,, \label{eq:tau6} 
\end{equation}
to ensure that the relative positions of the two paths $\Delta x(t=\tau_6)\approx0$~m, and the relative velocity $\Delta v(t=\tau_6)\approx 0$~ms$^{-1}$ (completion of interferometry). 
\begin{figure}
	\includegraphics[width=0.99\linewidth]{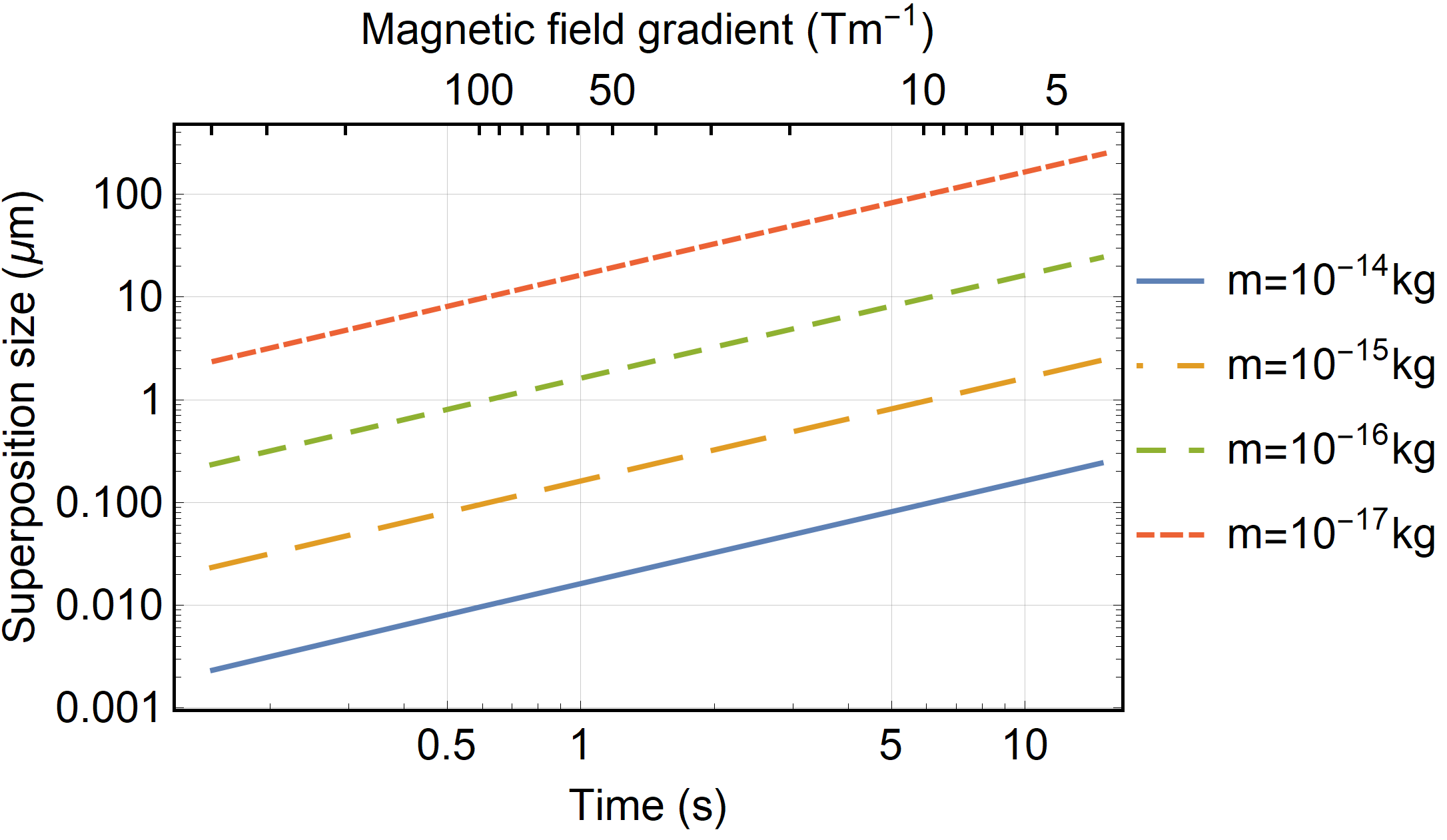}
	\caption{Superposition size vs. time. This shows the maximal superposition size achievable, as determined by the calculated paths. Note that for masses $m=10^{-17}$~kg and heavier these results are well approximated by Eq. \ref{eq:numerical sup size with time}.  \label{fig:superposition with time and mass}}
\end{figure}

This results in the superposition sizes as seen in Fig.\ref{fig:superposition with time and mass}, and the trajectories through the interferometer can be seen as in Fig.\ref{fig:Path figures}. Furthermore, the maximum superposition which occurs between $\tau_4< t<\tau_5$ can be determined numerically:
\begin{equation}
	\Delta x_{\textrm{max}} (m,\tau_6) \approx \left(\frac{1.6\times10^{-16}{\rm~Kg}}{m}\right)\left(\frac{\tau_6}{{\rm 1~sec}}\right)\times 10^{-6}\textrm{~m}
	 \label{eq:numerical sup size with time}
\end{equation}
for $B_0=10^{-2}$~T, $B_1=100\varepsilon$ and $m\gtrsim10^{-17}$~kg. The latter condition on the mass arises because smaller masses are subject to much smaller induced diamagnetic potential. For a lighter diamond, if we do not modify the times $\tau_i\left(\eta\right)$ and $B_1$, the masses would inevitably move through the zero-field region of the magnetic field, and therefore demands a different magnetic field setup, see~\cite{Folman2013, margalit2021realization}. Therefore, our analysis holds true for $m\geq 10^{-17}$~kg for $\chi_m\approx-6.2\times10^{-9}$~kgm$^{-3}$. Using Eq.\ref{eq:numerical sup size with time} we can estimate that achieving a superposition size of $20$~$\mu$m with a $10^{-17}$~kg mass requires a total time of $\tau_6\approx1.25$~s, which corresponds to the moderate magnetic field gradient $\eta\sim46.8$~Tm$^{-1}$, again using $B_0=10^{-2}$~T and $B_1=100\varepsilon$, which can be achieved in a laboratory~\cite{Folman2013, modena2012design}.

\section{Spin Phase Evolution \label{sec:Path phase difference}}

\begin{figure}
	\centering
	\includegraphics[width=0.95\linewidth]{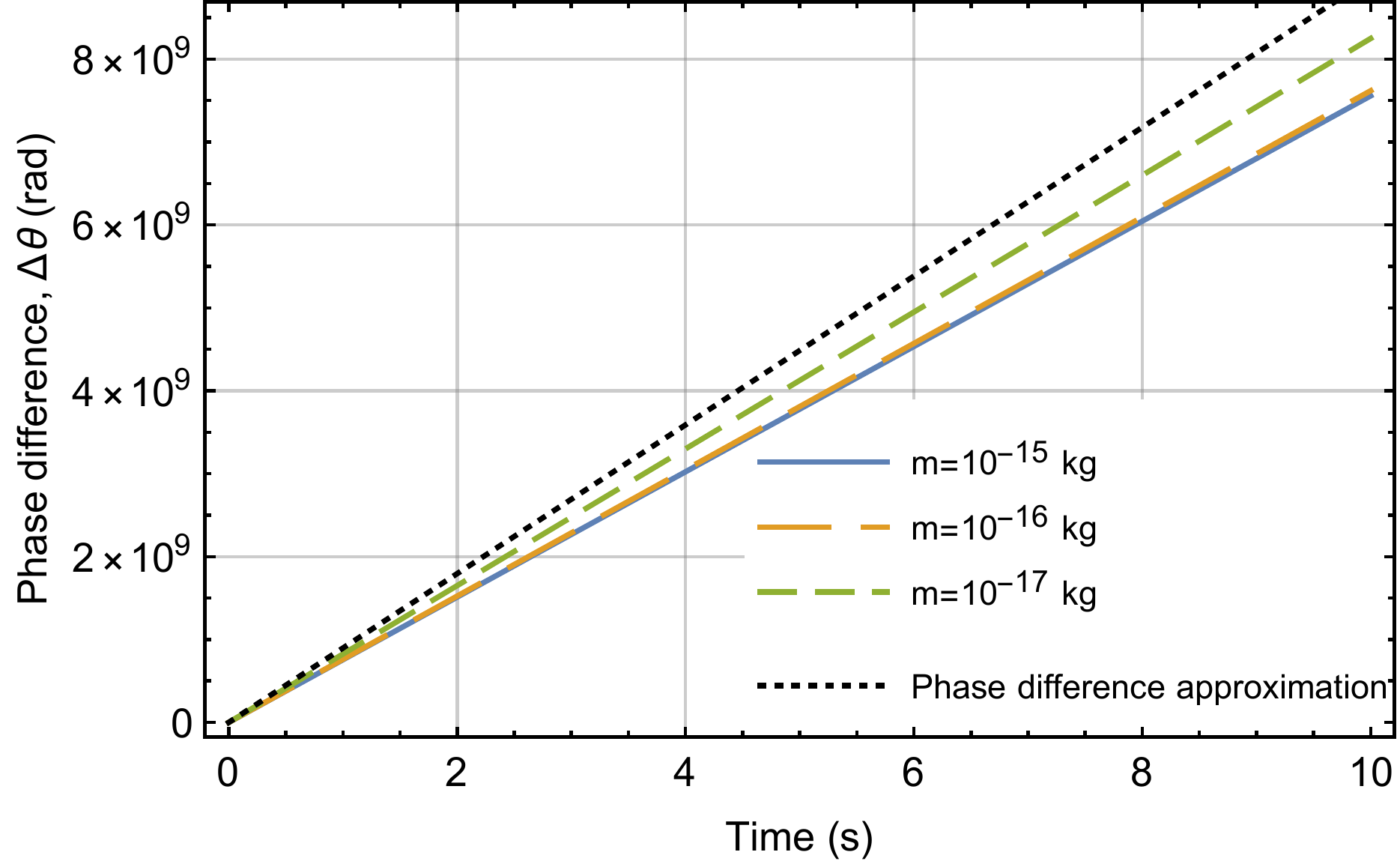}
	\caption{Phase difference magnitude scaling with total runtime for $m=10^{-15}$ kg, $m=10^{-16}$ kg, and $m=10^{-17}$ kg. This also shows the phase difference as predicted by Equation \ref{eq:Approx phase difference}. \label{fig:Phase difference}}
\end{figure}
To determine the output signal expected from such an interferometer it is necessary to consider the phase evolution difference between the two paths taken through the interferometer. The final output signal will be the result of a combination of the phase accumulated by the phase due to the path it takes through the interferometer, and external forces acting on the masses, such as gravitational or electromagnetic interactions which differ across the paths taken through the interferometer. For the moment however, we will neglect these outside sources due to highly implementation dependent nature of these. Given the entire Hamiltonian is used to determine the trajectories, it is sufficient to calculate the action for a \emph{free} spin travelling along a fixed trajectory given by the two trajectories through the interferometer. This will then yield a phase, $\theta$ accumulated along a path $\gamma$ of
\begin{equation}
	\theta=\frac{1}{\hbar}\int_{\gamma}\frac{p(t)^2}{2m} \textrm{d}t
\end{equation}
where $p(t)$ is the time dependent momentum of the particle. 
To evaluate this we can approximate the motion as being in three distinct segments: $\tau_0=0\le t\le \tau_1$, before the particle has reached the zero field region; $\tau_2< t\le \tau_3$, after the particle has reached the zero field region but before the spin has been reversed; and  $\tau_3< t\le \tau_4$, the remainder of the evolution, ending when the two wave packets are brought to overlap. Note that the time region $t\in\left[\tau_1,\tau_2\right]$ is extremely brief and does not significantly change the end result. In each of these sections the general particle momentum can be written as
\begin{equation}
	p^{+/-}(t)=-m\omega A_{i}^{+/-}\sin\left(\omega t+\phi_i^{+/-}\right)
\end{equation}
for the $+$ and $-$ arm of the interferometer and the value of $A^{+/-}$ and $\phi^{+/-}$ differs from one segment to the next. Thus the phase difference accumulated through any given stage of the interferometer will be
\begin{align}
	\Delta \theta=&\sum_{i\in\left\{1,3,4\right\}}\Bigg[\frac{m}{2\hbar}\int_{\tau_{i-1}}^{\tau_{i}}\left(\omega A_i^{+}\sin\left(\omega t+\phi_i^{+}\right)\right)^2 \nonumber\\
	&-\frac{m}{2\hbar}\int_{\tau_i}^{\tau_{i+1}}\left(\omega A_i^{-}\sin\left(\omega t+\phi_i^{-}\right)\right)^2 \Bigg]\nonumber\\
	=&\frac{m\omega}{8 \hbar}\left[2\omega\tau_1-\sin\left(2\omega\tau_1\right)\right]\left[\left(A_1^{+}\right)^2-\left(A_1^{-}\right)^2\right] \nonumber\\
	&+\sum_{i=3}^{4}\frac{m}{8\hbar}\Bigg[2\omega^2\left(T_{i}-T_{i-1}\right)\left(\left(A^{+}\right)^2-\left(A^{-}\right)^2\right) \nonumber\\
	&+\sin\left(2\phi^{+}+2T_{i-1}\omega\right)-\sin\left(2\phi^{-}+2T_{i-1}\omega\right) \nonumber\\
	&-\sin\left(2\phi^{+}+2T_{i}\omega\right)+\sin\left(2\phi^{-}+2T_{i}\omega\right)\Bigg] \label{eq:Full phase difference}
\end{align}
Now it is suitably insightful to consider in detail the phase difference accrued when $t\in\left[\tau_0,\tau_1\right]$, noting that $\sin\left(2\omega\tau_1\right)\approx0$:
\begin{align}
	\Delta \theta\left(t=\tau_1\right)\approx&\frac{m}{8\hbar}\Bigg[2\omega^2\tau_1\left(\left(A_{1}^{+}\right)^2-\left(A_1^{-}\right)^2\right) \nonumber\\
	&-\sin\left(2\tau_1\omega\right)+\sin\left(2\tau_1\omega\right)\Bigg] \nonumber\\
	=&\frac{m}{8\hbar}2\omega^2\tau_1\left[\left(\frac{B_0}{\eta}+\alpha\right)^2-\left(\frac{B_0}{\eta}-\alpha\right)^2\right] \nonumber\\
	=&\frac{geB_0}{2 m_e}\tau_1 \nonumber\\
\end{align}
Given that in each time segment considered here, the particles are simply acting as harmonic oscillators, flipping between two different harmonic wells, the entire path phase difference can be approximated as
\begin{equation}
	\Delta\theta\approx\frac{geB_0}{2 m_e}\left(\frac{\tau_1}{\tau_4}-\frac{\tau_3-\tau_1}{\tau_4}+\frac{\tau_4-\tau_3}{\tau_4}\right)\tau_4 \label{eq:Approx phase difference}
\end{equation}
where the extra terms in equation \ref{eq:Full phase difference} are neglected and the negative time scaling occurs when the internal spin direction is reversed relative to the external field. Note that this occurs by reversing the field magnitude, not the particle spin state. This entire term $\left(\frac{\tau_1}{\tau_4}-\frac{\tau_3-\tau_1}{\tau_4}+\frac{\tau_4-\tau_3}{\tau_4}\right)\approx0.5$ can be viewed as a comparison between the time spent with the masses being accelerated away from one another and time spent being attracted towards one another. This would typically be equal and as such this entire phase would not usually appear. It is only due to the asymmetry in this interferometer set up that it occurs. Thus, for the full trajectories, the path phase difference will be $\Delta \theta	\propto \tau_4$. This can be seen clearly in Figure \ref{fig:Phase difference}, which shows the phase scaling linearly with time and almost independently of mass. This also shows the approximate phase difference given by Equation \ref{eq:Approx phase difference} and how it compares to the exact values. This also allows for the required stability to be estimated, using equation \ref{eq:Approx phase difference}:  $\Delta\theta\approx\frac{ge}{4m_e}B_0\tau_4\sim B_0\tau_4 \times10^{11}\textrm{ T}^{-1}\textrm{s}^{-1}$; to keep the final phase uncertainty $\delta\left(\Delta\theta\right)<1$ will require
\begin{align}
	10^{-11}\textrm{ Ts}>&B_0\times\delta t \\
	10^{-11}\textrm{ Ts}>&\delta B_0\times t
\end{align}
which places an strict requirement on both timing certainty, $\delta t$; and bias field stability, $\delta B_0$. This will serve as a further challenge that must be met before such an experiment can be fully realised.

\section{Discussion}
 In this paper, we have provided a simple mechanism to accelerate heavy neutral masses, with embedded spin, so as to create large spatial superpositions. Using Eq.(\ref{eq:numerical sup size with time}) we estimate that one can create, for example, the spatial superposition of $\sim 20{\rm~\mu m}$ for masses, as heavy as $10^{-17}$~kg. Our simple scheme fills the gap required for a realistic ``wavefunction splitting'' of large masses to achieve a large spatial superposition, and herewith opens up new vistas for testing and probing both the classical and quantum nature of gravity while also giving access to unprecedented sensing opportunities. 

The time duration of coherence, and hence the experiment, is the {\em only} limiting factor, but (a) spin coherence times are perpetually rising (approaching $1$s \cite{bar2013solid, abobeih2018one}, even 30 s \cite{muhonen2014storing, farfurnik2015optimizing})-- adapting these to nanocrystals remains an open problem,  but is not fundamentally restricted \cite{knowles2014observing}, (b) spatial coherence times can be made $\sim 100$ s \cite{Bose:2017nin,vandeKamp:2020rqh,MIMAC2018,Toros:2020dbf}, by challenging, but achievable  pressures \cite{fitzakerley2016electron}, temperatures \cite{marx2014dry}, distances from other sources and fluctuations.  For example, a decoherence rate below $0.1$ Hz is achievable for diamond spheres of masses $\sim10^{-14}$ kg. This is expected \cite{schut2021improving} for internal temperatures of $0.15$ K,  an environmental temperature of $1$ K and a environmental gas number density of $10^8$ m$^{-3}$.

 Effort will also have to be taken to initialise the particle with the spin direction aligned with the external applied magnetic field to minimise spurious torques arising from the spin-magnetic field coupling.  This can be accomplished already before being released from a trap,  for example,  by using anisotropically shaped nanoparticles,  which can be aligned with any given direction in space by using linearly polarized lasers or electric fields  \cite{stickler2021quantum} or magnetic fields~\cite{perdriat2021angle}.  There will still be harmonic motion about this orientation axis (called librations),  which generally have much higher frequencies \cite{liu2017coupling} than typical centre of mass trapping frequencies in tweezers.  As the latter has already been cooled to the ground state by feedback cooling \cite{tebbenjohanns2021quantum, delic2020cooling, tebbenjohanns2020motional, magrini2021real},  one would expect the former (librations) to be possible  to be cooled to the ground state also.  In fact,  these librations have been substantially cooled~\cite{delord2020spin}.  Once it is cooled close to the ground state,  the spin direction will be effectively exactly aligned to the external magnetic field.  If the magnetic field is maintained throughout the interference in the same direction then the spin, and hence the nanoparticle, receives no torque from it.  We also note that it has already been shown that the internal degrees of freedom (phonons) do not pose a practical problem~\cite{henkel2021internal}.
We have shown that there are also requirements on the magnetic field fluctuations and timing certainty to ensure a stable interference signal, both at achievable levels, however \cite{ruster2016long, gohil2020measurements, sochnikov2020femto, mourou1981picosecondswitch}. To conclude, we have found an explicit scheme to create large spatial superpositions, with the size increasing in proportion to achievable coherence times, but otherwise not limited.  

{\it Acknowledgements}:
R. J. M. is supported by a  EPSRC DTP departmental studentship at UCL.  AM's research is funded by the Netherlands Organisation for Science and Research (NWO) grant number 680-91-119. R. F. is supported by the Israel Science Foundation. SB would like to acknowledge EPSRC Grant Nos. EP/N031105/1 and  EP/S000267/1.

\bibliography{bibliography}

\appendix
\section{Solving the particle dynamics \label{sec:solving the dynamics}}

\begin{figure}
	\centering
	\includegraphics[width=0.95\linewidth]{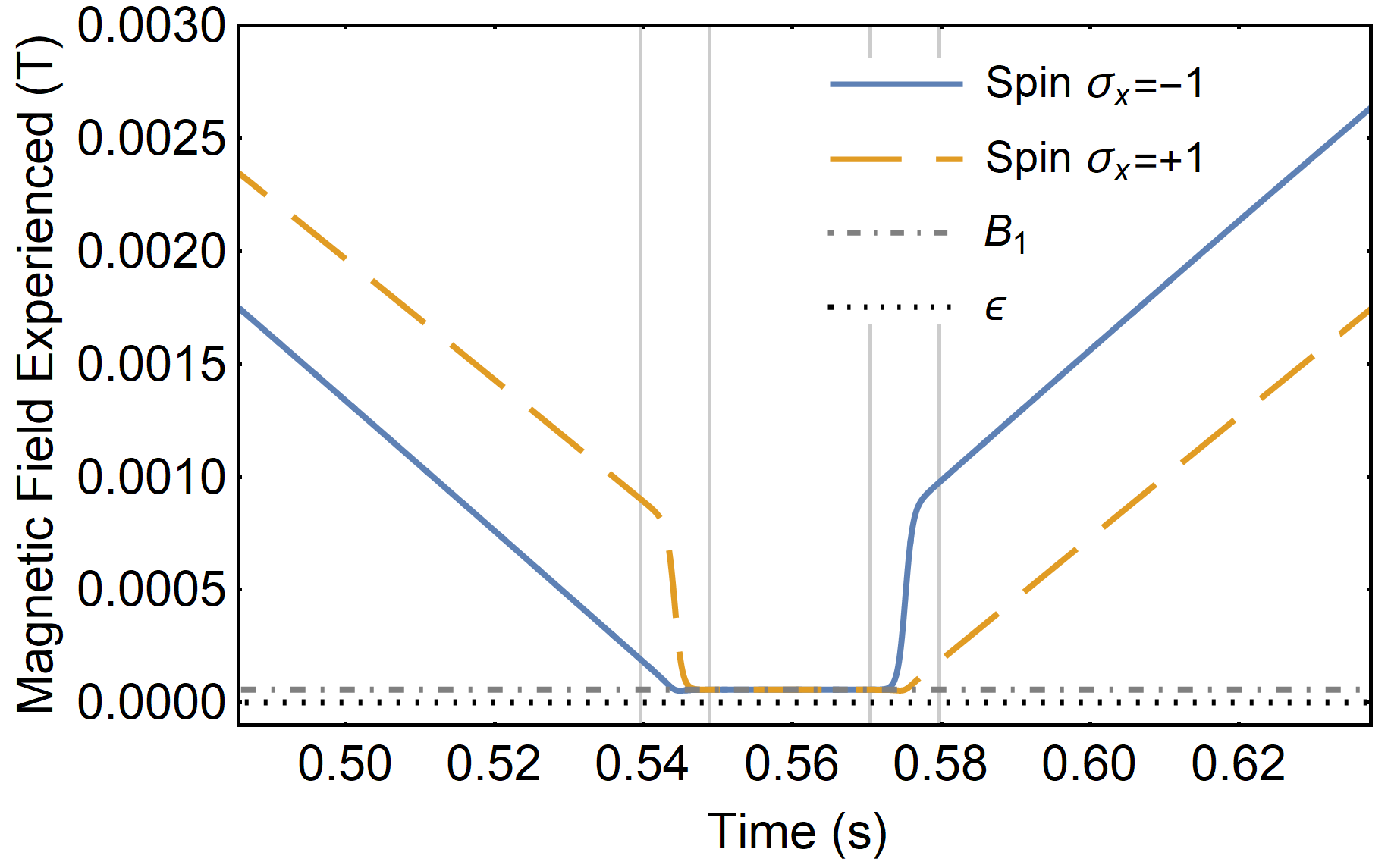}
	\caption{ Magnetic field experienced by each arm of the interferometer with time. The vertical grey lines represent times $\tau_i$ for $i={1,2,3,4}$ using $B_0=10^{-2}$~T, $\eta=40$~Tm$^{-1}$ and $B_1=100\varepsilon$~T. \label{fig:Experienced Bfield}}
\end{figure}

In this supplementary material, we provide our analytical and numerical treatment of the equations of motion. The magnetic fields used to create the superposition are
\begin{align}
	\vec{B}\left(x,y\right)=&\left(B_0-\eta x\right)\hat{x}+\eta y\hat{y} \label{eq:Bfield1 appendix}\\
	\vec{B}\left(x,y\right)=&B_1\hat{x} \label{eq:Bfield2 appendix}\\
	\vec{B}\left(x,y\right)=&-\left(B_0-\eta x\right)\hat{x}-\eta y\hat{y} \label{eq:Bfield3 appendix}
\end{align}
with the magnetic field being mapped between Eq.~\ref{eq:Bfield1 appendix} to Eq.~\ref{eq:Bfield2 appendix} when $t\in\left[\tau_1,\tau_2\right]$, and between Eq.~\ref{eq:Bfield2 appendix} to Eq.~\ref{eq:Bfield3 appendix} when $t\in\left[\tau_3,\tau_4\right]$. Figure \ref{fig:Experienced Bfield} shows the detail of the experienced magnetic field for each arm of the interferometer during the period in which the magnetic field is modified. This shows that provided the value chosen for $B_1$ is sufficiently large that the minimum allowable experienced magnetic field $\varepsilon$ can be avoided while ensuring a smooth and adiabatic magnetic field transition. When not in this phase ($t\notin\left[\tau_1,\tau_4\right]$) the motion is simply governed by the harmonic oscillator potential, whose solution is given by:
\begin{equation}
	x(t)=A \cos\left(\omega t + \phi\right) + C(s_{x}) \label{eq:general motion Appendix}
\end{equation}
where $A$ and $\phi$ are determined by the initial conditions, $\omega=\sqrt{-\frac{\chi_m}{\mu_0}}\eta$ is the frequency of the diamagnetic trap and $C(s_{x})$ is determined by the spin states and magnetic field (Eqs. \ref{eq:Bfield1 appendix} and \ref{eq:Bfield3 appendix}). Note that we are always within the adiabatic limit throughout these times, such that $\dot \omega /\omega^2 <1$. We also have 
\begin{equation}
	C^{s_{x}=\pm1}= {B_0}/{\eta}\pm\alpha \label{eq:electronic spin solution appendix}\,,
\end{equation}
where $\alpha=-\mu_{\pm}\frac{\mu_{0}}{\chi_{m} m \tilde{\eta}(t)}=-\frac{g e \hbar}{2 m_e}\frac{\mu_{0}}{\chi_{m} m \tilde{\eta}(t)}$. 
The trajectory through these times, $t\notin\left[\tau_1,\tau_4\right]$, can be constructed by simply assembling the solutions piecewise. To do this, the values of the constants are given by the function of the initial conditions at each time segment. Specifically, if $x(t_0)=x_0$ and $\dot{x}(t_0)=v_0$ serve as the initial conditions, then
\begin{eqnarray}
	A=&-\left(C-x_0\right)\left(1+\frac{v_0^2}{\omega^2}\left(C-x_0\right)^{-2}\right)^{1/2} \\
	\phi=&\tan^{-1}\left(\frac{v_0}{\omega\left(C-x_0\right)}\right)-\omega t_0.
\end{eqnarray}
For example, if we consider the particle to be initialised in a superposition of spin states $\left|+1\right\rangle$ and $\left|-1\right\rangle$ at the origin with zero initial velocity, then we can define the initial motion of the two arms as:
\begin{eqnarray}
	x_1^{+}(t)=&-\left(\frac{B_0}{\eta}+\alpha\right) \cos\left(\omega t\right) + \left(\frac{B_0}{\eta}+\alpha\right)\\
	x_1^{-}(t)=&-\left(\frac{B_0}{\eta}-\alpha\right) \cos\left(\omega t\right) + \left(\frac{B_0}{\eta}-\alpha\right)\,.
\end{eqnarray}
When the magnetic field is given by Eq.~\ref{eq:Bfield2 appendix} there will be no spin dependant acceleration, that is, there is no force acting to either create or destroy the spatial superposition. Thus we want to minimise $\tau_4-\tau_1$. To do this, the initial evolution, $x_1(t)$, should be maintained for as long as possible while still ensuring that 
\begin{equation}
	x\notin\left[\frac{B_x(0,0)-\varepsilon}{\partial B},\frac{B_x(0,0)+\varepsilon}{\partial B}\right]\,, \label{eq:limit on x - appendix}
\end{equation} 
is satisfied. Only as the particle approaches the boundary set by Eq.\ref{eq:limit on x - appendix} is the magnetic field is modified (this marks the time $\tau_1$). Specifically, this is done as the magnetic field experienced by the forward most trajectory approaches $B_1$:
\begin{equation}
	x_1^{+}(t\approx\tau_1)=\frac{B_0-B_1}{\eta}\,,
\end{equation}
where we write $x_1^{-}(\tau_1)=A^{-}_1 \cos\left(\omega\tau_1\right)+c^{-}_1$. 

Similarly, the non-zero magnetic field gradient should be returned as soon as possible while ensuring Eq. \ref{eq:limit on x - appendix} holds. As such, the magnetic field gradient should begin being restored as the magnetic field experienced by the rearmost trajectory ($x_2^{-}(t)$) approximately the magnetic field as given by Eq.~\ref{eq:Bfield3 appendix}, that is:
\begin{align}
	B_1=&-B_0+\eta x^{-}_2\left(\tau_3\right) \nonumber\\
	x^{-}_2\left(t\approx\tau_3\right)=&\frac{B_1-B_0}{\eta}
\end{align}
where $x^{+}_2$ and $x^{-}_2$ are the trajectories when $t\in\left[\tau_1,\tau_4\right]$, and are found by numerically integrating the equations of motion.

The final two stages of the trajectories are given by:
\begin{align}
	x_3^{+}(t)=&A^{+}_{3} \cos\left(\omega t + \phi^{+}_{3}\right) + C^{+}_{3} \label{eq:x3+}\,,\\
	x_3^{-}(t)=&A^{-}_{3} \cos\left(\omega t + \phi^{-}_{3}\right) + C^{-}_{3} \label{eq:x3-}\,,\\
	x_4^{+}(t)=&A^{+}_{4} \cos\left(\omega t + \phi^{+}_{4}\right) + C^{+}_{4} \label{eq:x4+}\,,\\
	x_4^{-}(t)=&A^{-}_{4} \cos\left(\omega t + \phi^{-}_{4}\right) + C^{-}_{4} \label{eq:x4-}\,,
\end{align}
where
\begin{widetext}
	\begin{eqnarray}
		A^{+}_3=-&\left(C^{+}_3-x^{+}_2(\tau_2)\right)\left(1+\frac{\left(v^{+}_2(\tau_2)\right)^2}{\omega^2}\left(C^{+}_3-x^{+}_2(\tau_2)\right)^{-2}\right)^{1/2}\,, \nonumber \\
		A^{-}_3=-&\left(C^{-}_3-x^{-}_2(\tau_2)\right)\left(1+\frac{\left(v^{-}_2(\tau_2)\right)^2}{\omega^2}\left(C^{-}_3-x^{-}_2(\tau_2)\right)^{-2}\right)^{1/2}\,,\nonumber \\
		\phi^{+}_3=&\tan^{-1}\left(\frac{v^{+}_2(\tau_2)}{\omega\left(C^{+}_3-x^{+}_2(\tau_2)\right)}\right)-\omega \tau_2\,, 
		\nonumber\\
		\phi^{-}_3=&\tan^{-1}\left(\frac{v^{-}_2(\tau_2)}{\omega\left(C^{+}_3-x^{-}_2(\tau_2)\right)}\right)-\omega \tau_2\,,~~\nonumber \\ 
		C^{+}_3=& -\left(\frac{B_0}{\eta}-\alpha\right)\,, \nonumber\\
		C^{-}_3=&-\left(\frac{B_0}{\eta}+\alpha\right)\,,
	\end{eqnarray}
	and
	\begin{eqnarray}
		A^{+}_4=&-\left(C^{+}_4-x^{+}_3(\tau_3)\right)\left(1+\frac{\left(v^{+}_3(\tau_3)\right)^2}{\omega^2}\left(C^{+}_4-x^{+}_3(\tau_3)\right)^{-2}\right)^{1/2}\,,\nonumber \\
		A^{-}_4=&-\left(C^{-}_4-x^{-}_3(\tau_3)\right)\left(1+\frac{\left(v^{-}_3(\tau_3)\right)^2}{\omega^2}\left(C^{-}_3-x^{-}_3(\tau_3)\right)^{-2}\right)^{1/2}\,,\nonumber \\
		\phi^{+}_4=&\tan^{-1}\left(\frac{v^{+}_3(\tau_3)}{\omega\left(C^{+}_4-x^{+}_3(\tau_3)\right)}\right)-\omega \tau_3\,, 
		\nonumber \\
		\phi^{-}_4=&\tan^{-1}\left(\frac{v^{-}_3(\tau_3)}{\omega\left(C^{+}_4-x^{-}_3(\tau_3)\right)}\right)-\omega \tau_3\,, 
		\nonumber \\
		C^{+}_4=& - \left(\frac{B_0}{\eta}+\alpha\right) \,,\nonumber \\
		C^{-}_4=& - \left(\frac{B_0}{\eta}-\alpha\right).
	\end{eqnarray}
\end{widetext}
The values for the times $\tau_5$ and $\tau_6$ are fixed by the following conditions:
\begin{enumerate}
	\item $\Delta x(\tau_6)=x_4^{+}(\tau_6)-x_4^{-}(\tau_6)=0$ and $\Delta v(\tau_6)=v_4^{+}(\tau_6)-v_4^{-}(\tau_6)=0$, such that the two arms of the interferometers are brought together to overlap in both the position and the momentum space, respectively.
	\item $x_4^{+}(\tau_6)>\frac{B_0+\varepsilon}{\eta}$, such that the Majorana spin flip region of the magnetic field is again avoided.
\end{enumerate}
Note that the appropriate time scales which met the above conditions (conditions 1. and 2.) were found {\it solely} numerically. This has lead to a set of times and the corresponding magnetic field gradients applied during the time period, which we have tabulated in the table \ref{tab:tau values}.
\begin{table}
	\centering
	\caption{$\tau_i$ values used to calculate wave-packet trajectories for example magnetic field gradients ($\eta$) \label{tab:tau values}}
	\begin{tabular}{|c|c|c|c|c|}
		\hline
		& $\eta=4$~Tm$^{-1}$ & $\eta=40$~Tm$^{-1}$ & $\eta=400$~Tm$^{-1}$ \\
		\hline
		$\tau_1$ & 5.39~s & 0.534~s & 0.0493~s \\
		\hline
		$\tau_2$ & 5.39~s & 0.539~s & 0.0539~s \\
		\hline
		$\tau_3$ & 5.80~s & 0.580~s & 0.0580~s \\
		\hline
		$\tau_4$ & 5.80~s & 0.584~s & 0.0626~s \\
		\hline
		$\tau_5$ & 9.01~s & 0.902~s & 0.0913~s \\
		\hline
		$\tau_6$ & 14.8~s & 1.48~s & 0.148~s \\
		\hline
		
	\end{tabular}
\end{table}

There are a couple of points to note here; the time $\tau_6$, set such that $\Delta v(\tau_6)=0$ will automatically minimise $\Delta x(\tau_6)$. Also, the value of $\Delta x$ evaluated at time $\tau_6$ is continuous in $\tau_5$, and there exists times $t_1$ and $t_2\in\mathbb{R}$ such that, when $\tau_5=t_1~\Delta x(\tau_6)>0$ and when $\tau_5=t_2~\Delta x(\tau_6)<0$. In a nut-shell the procedure to find the correct values of $\tau_5$ and $\tau_6$ are:
\begin{enumerate}
	\item To make an initial guess for the value of $\tau_5$ which was used to calculate a complete trajectory, typically this is 
	$\tau_5=2\tau_4$.
	\item From this we have evaluated the value of $\tau_6$, which we have determined using the relation $\Delta v(\tau_6)=0$, and
	\item the corresponding value of $\Delta x(\tau_6)$ was then evaluated. This allows the the assumed value for $\tau_5$ to be optimised accordingly. Specifically noting that increasing $\tau_5$ will lead to decrease the value of $\Delta x(\tau_6)$.	
\end{enumerate}

\end{document}